\documentclass[ superscriptaddress,reprint,amsmath,amssymb,aps,pra,longbibliography]{revtex4-1}
\usepackage{dcolumn} 

\setcitestyle{numbers,super}
\usepackage{fnpct}
\usepackage{placeins} 
\usepackage{upgreek}
\usepackage{epstopdf} \usepackage{amsmath} \usepackage{physics} \usepackage{graphicx} 
 \usepackage{bm} \usepackage[margin=0.7in]{geometry}

\begin{document}
\preprint{APS/123-QED}

\title{
Quantum Reference Beacon-Guided Super-Resolution Optical Focusing in Complex Media
}
\author{
Donggyu Kim
}
\email{donggyu@mit.edu}
\affiliation{Department of Mechanical Engineering, Massachusetts Institute of Technology, 77 Massachusetts Avenue, Cambridge, MA, 02139, USA}
\affiliation{Research Laboratory of Electronics (RLE), Massachusetts Institute of Technology, \\50 Vassar Street, Cambridge, MA, 02139, USA}

\author{
Dirk R. Englund
}
\affiliation{Research Laboratory of Electronics (RLE), Massachusetts Institute of Technology, \\50 Vassar Street, Cambridge, MA, 02139, USA}
\affiliation{Department of Electrical Engineering and Computer Science, Massachusetts Institute of Technology, 50 Vassar Street, Cambridge, MA, 02139, USA}


\begin{abstract} 
Optical random scattering is generally considered to be a nuisance of microscopy that limits imaging depth and spatial resolution. Wavefront shaping techniques have recently enabled optical imaging at unprecedented depth, but a remaining problem is also to attain super-resolution within complex media. To address this challenge, we introduce a new technique to focus inside of complex media by the use of a quantum reference beacon (QRB), consisting of solid-state quantum emitters with spin-dependent fluorescence. This QRB provides subwavelength guidestar feedback for wavefront shaping to achieve an optical focus below the microscope's diffraction limit. We implement the QRB-guided imaging approach using nitrogen-vacancy centers  in diamond nanocrystals, which enable optical focusing with a subdiffraction resolution below 186 nm ($\approx \lambda/3.5\mbox{NA}$), where the microscope's NA=0.8. This QRB-assisted wavefront shaping paves the way for a range of new applications, including deep-tissue quantum enhanced sensing and individual optical excitation of magnetically-coupled spin ensembles for applications in quantum information processing.
\end{abstract} 

\maketitle
Optical random scattering in complex media, such as biological tissues, distorts an incident optical focus, reducing the resolution and imaging depth of optical microscopy. However, it was recently shown that random scattering does not lead to the permanent loss of focusing capability; instead it randomizes the incident focus in a deterministic way. By reversing this scattering, it becomes possible to focus\cite{vellekoop2010exploiting,van2011scattering,park2013subwavelength} and even to image\cite{Popoff:2010ac, choi2011overcoming,park2014full} through complex media. Moreover, random scattering can actually benefit\cite{vellekoop2010exploiting, van2011scattering, choi2011overcoming, park2013subwavelength,park2014full} microscopy by permitting a spatial resolution below the diffraction-limit of $\lambda/ 2\mbox{NA}$, where NA is the numerical aperture of the microscope objective. This super-resolution is possible because random scattering couples optical modes with high in-plane momentum from the sample to the microscope objective, much like a disordered grating. By extending this principle to evanescent modes of the sample, far-field superlenses for near-field focusing \cite{park2013subwavelength} and imaging \cite{park2014full} have been achieved. 

Reversing random scattering requires feedback from the target focal points. In particular, focusing light inside of complex media requires a type of ``guidestar (GS)''  that provides feedback of the interior optical field\cite{horstmeyer2015guidestar}. This feedback guides incident wavefront adjustments to focus the scattered light into the GS point. In the last decade, various forms of GSs have been implemented, including fluorescence \cite{vellekoop2008demixing}, ultrasound \cite{xu2011time, si2012fluorescence, wang2012deep, judkewitz2013speckle, chaigne2014controlling, lai2015photoacoustically}, nonlinear reference beacons \cite{katz2014noninvasive}, and kinetic objects \cite{ma2014time, ruan2015optical}. However, the spatial resolution using these types of GSs has been far from the super-resolution limit\cite{horstmeyer2015guidestar}. To push this resolution to or below the diffraction limit requires two key advances: (i) the physical size of the GS needs to be of subwavelength scale, and (ii) it must be possible to resolve subdiffraction features of randomly scattered light\cite{park2013subwavelength,park2014full}. A subwavelength aperture used in scanning near-field optical microscopy (SNOM) satisfies these conditions, but this technique does not permit imaging within a complex medium. To address these challenges, we introduce quantum reference beacons (QRBs).

The QRB we propose consists of solid-state quantum emitters with spin-dependent fluorescence. An example is the nitrogen vacancy (NV) center in diamond, which has emerged as a leading quantum system for quantum sensing \cite{balasubramanian2008nanoscale, maze2008nanoscale, kucsko2013nanometer} and quantum information processing \cite{jiang2009repetitive, neumann2010single, Robledo:2011aa, bernien2013heralded, taminiau2014universal, waldherr2014quantum, kalb2017entanglement}. By resonantly driving electron spin transitions of each QRB, the spin-dependent fluorescence produces the subwavelength GS feedback that enables super-resolution focusing within complex media. We demonstrate our proposal with ensembles of NV centers in subwavelength diamond nanocrystals, and show super-resolution focusing inside of a disordered scattering medium with a resolution below 186 nm ($\approx \lambda/ 3.5\mbox{NA}$ with $\lambda = 532 \text{ nm}$). 

\begin{figure*} [!hbt]
\centering
\includegraphics[scale=0.42]{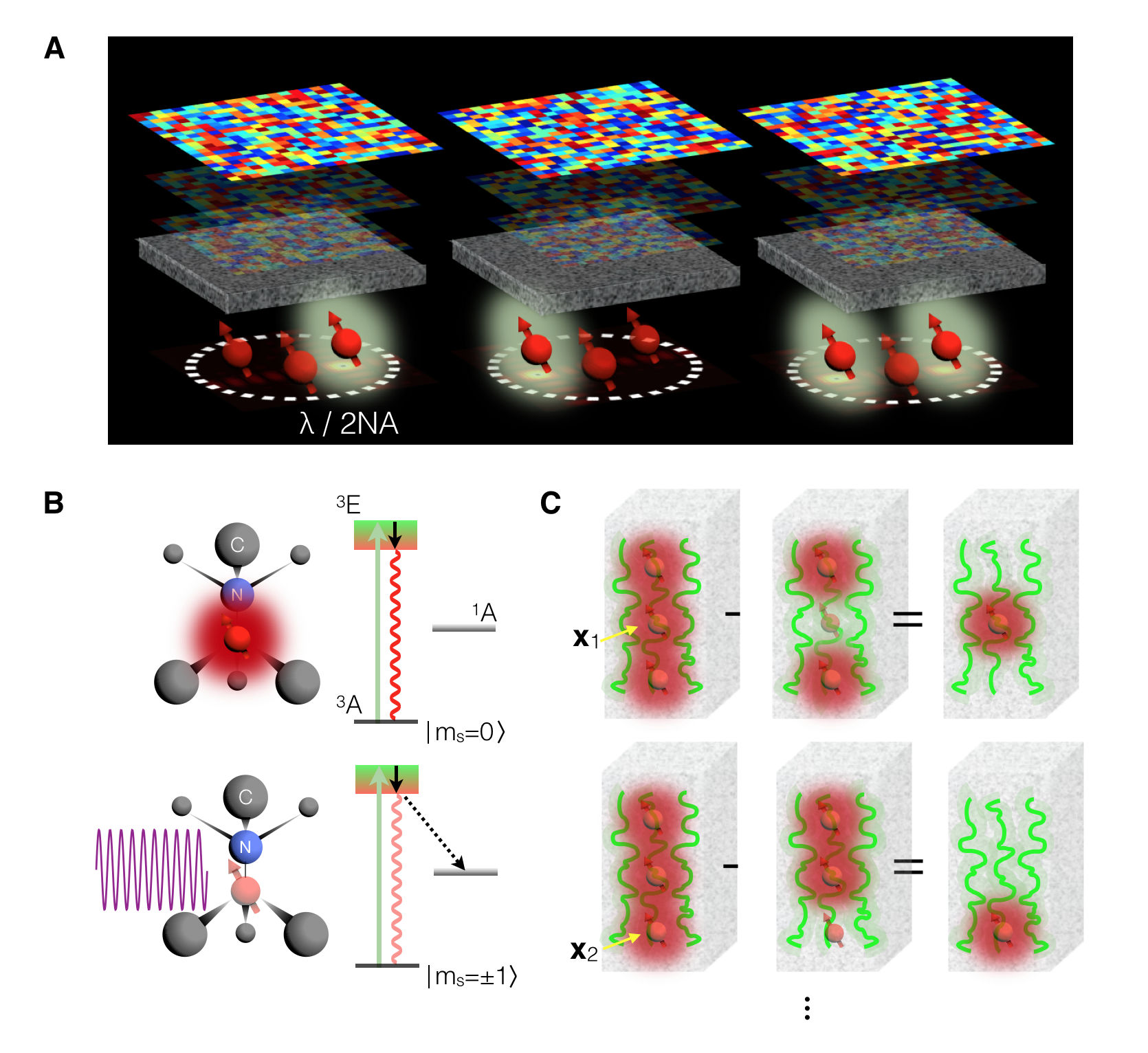}
\caption{
\textbf{Wavefront shaping guided by quantum reference beacons (QRBs)}
(\textbf{A}) Optical random scattering in complex media distorts the incident optical field. However, this distortion can be reversed by shaping the incident wavefront. Embedded QRBs provide feedback about subwavelength features of the scattered optical fields, guiding the wavefront shaping process. This approach enables, for example, super-resolution focusing deep inside of complex media or individual spin-qubit measurement in a diffraction-limited area (the dashed circle). 
(\textbf{B}) Nitrogen-vacancy (NV) centers in diamond with spin-dependent fluorescence: Electrons with the spin magnetic sublevels $\ket{m_s = \pm 1}$ preferentially decay (dashed black arrow) to the dark metastable state ($^1$A), once they are optically pumped to the excited states $^3$E (green arrow), resulting in reduced fluorescence than that from the sublevel $\ket{m_s = 0}$. This spin-dependent fluorescence enables optically detectable magnetic resonance (ODMR). 
(\textbf{C}) The QRB-GS feedback is produced with the spin-dependent fluorescence: To measure the optical field on the QRB positioned at $\mathbf{x}_1$, its fluorescence is selectively reduced by electron spin resonance (ESR). The change of collected fluorescence determines the optical field at $\mathbf{x}_1$. This process can be repeated for another position at $\mathbf{x}_2$ as shown in the bottom plot.
}
\end{figure*}

\begin{figure*} [!hbt]
\centering
\includegraphics[scale=0.42]{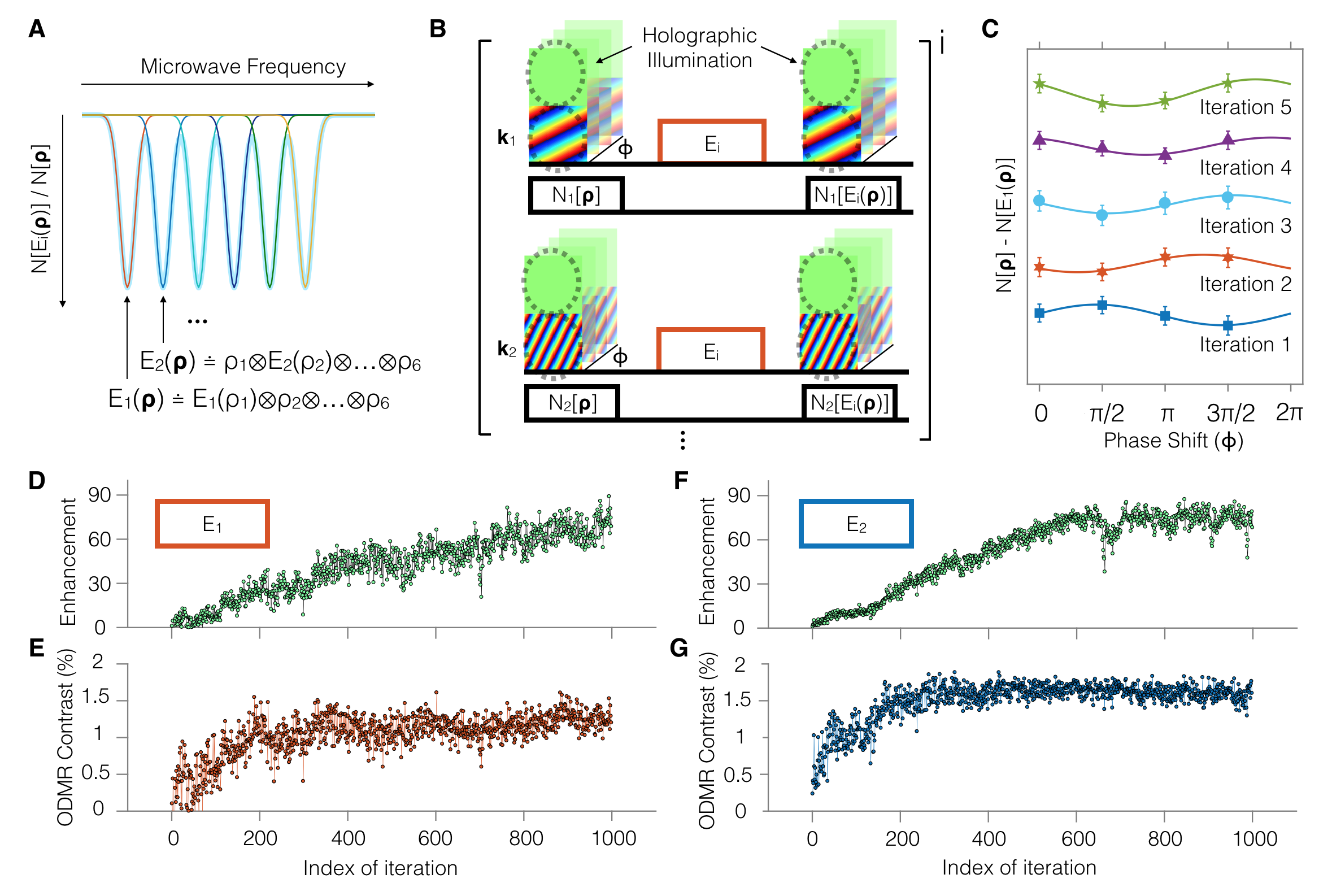}
\caption{
\textbf{Iterative wavefront optimization with QRB-GS feedback.}
(\textbf{A}) $\{ \rho_i \}$ label the electron spin states of QRBs, and an external magnetic field gradient splits their individual resonance frequencies. Quantum operators $\{ E_i \}$ drive the electron spin transition of target QRBs. 
(\textbf{B}) Measurement sequences for the iterative wavefront optimization: the Fourier basis modes of the incident wavefront ($\mathbf{k}_1$, $\mathbf{k}_2$, ...) are encoded into holographic illuminations, in which the basis modes interfere with the reference plane wave for complex field readout (See Supplementary text S4 and S5 for details). The overall fluorescence difference with $\{ E_i \}$ (i.e. $N_j[\pmb{\rho}] - N_j[E_i(\pmb{\rho})]$) produces the QRB-GS feedback $S_{i,j}$. $\phi$ describes the phase of each basis mode relative to the reference plane wave. 
(\textbf{C}) Modulation of the QRB-GS feedback in the iterative wavefront optimization: In each step, the phase $\phi$ of the basis modes is adjusted to compensate for the phase offset of the modulation. 
(\textbf{D}) and (\textbf{F}) The iterative wavefront optimization with the QRB-GS feedback: Two QRBs have the electron spin resonance frequencies at $\nu_1 = 2.825 \text{ GHz}$ and $\nu_2 = 2.762 \text{ GHz}$. The resonant microwaves continuously drive the resonances to produce the QRB-GS feedback, so that the  incident optical fields can be iteratively updated to optimize the QRB-GS feedback signal strength. 
(\textbf{E}) and (\textbf{G}) The ODMR contrast at $\nu_1$ and $\nu_2$ for the iterative optimization processes.
}
\end{figure*}

Figure 1 illustrates the approach to QRB-guided wavefront shaping in microscopy. A wavefront shaper adjusts basis modes (shown as individual pixels in Fig. 1A) of the incident wavefront to interfere scattered light constructively at target GS points. This specific wavefront adjustment is determined from the QRB-GS feedback. This feedback signal is created by applying a magnetic field gradient across the sample so that one of several QRBs inside a diffraction-limited volume can be selectively driven into its dark magnetic sublevels, as indicated in Fig. 1C and detailed below. 

Specifically, the QRB-GS feedback signal is needed to measure the transmission matrix\cite{horstmeyer2015guidestar} that characterizes the light propagation through a complex medium (See Supplementary text S1 and S2 for details). We label the electron spin state of the embedded QRBs at $\{ \mathbf{x}_i \} = \mathbf{x}_1,  \cdots, \mathbf{x}_N$ with a spin density operator $\pmb{\rho} = \rho_1 \otimes \rho_2 \otimes \cdots \otimes \rho_N$. An external magnetic field gradient separates their resonance frequencies $\{\nu_i \}$ by the Zeeman effect. In principle, $\{\mathbf{x}_i \}$ could then be reconstructed from $\{\nu_i \}$ and knowledge of the external magnetic field gradient. Resonant driving of each $\{\rho_i\}$ spin transition is represented through a quantum operator $\{ E_i \}$. When the $j$th incident basis mode is coupled into the medium, the QRB-GS feedback $S_{i,j}$ for $\mathbf{x}_i$ is described by
\begin{equation}
S_{i,j} = N_j[\pmb{\rho}] - N_j[E_i (\pmb{\rho})] = \abs{t_{i,j}}^2 \Delta \sigma_i \Delta \gamma.
\end{equation}
Here, $N_j[\pmb{\rho}]$ and $N_j[E_i (\pmb{\rho})]$ denote the fluorescence photon numbers collected for unit integration, $t_{i,j}$ is the transmission matrix element (i.e. the scattered optical field at $\mathbf{x}_i$ for the $j$th incident basis mode), $\Delta\sigma_i = \frac{1}{2} \text{tr}[ \sigma_z \{ \rho_i - E_i (\rho_i) \}]$ where $\sigma_z$ is the Pauli-$z$ operator, and $\Delta \gamma$ represents the variance of the collected spin-dependent fluorescence between the optically bright and dark spin states (Fig. 1B). Figure 2 summarizes the iterative wavefront adjustments due to the QRB-GS feedback.  

The spatial resolution of our method is determined by the ESR lineshape (See Supplementary text S2 for details), since  the lineshape sets the point spread function (PSF) of the QRB-GS feedback that confines $\{ E_i \}$ only to the target QRBs (Fig. 2A). Specifically, a magnetic field gradient $dB/dx$ translates the (mean) resonance linewidth $\delta\nu$ to the spatial resolution $\Delta d_\text{QRB}$ of the effective PSF: 
\begin{equation}
\Delta d_\text{QRB} =  \frac{ \delta\nu}{\gamma_e (dB/dx)}
\end{equation}
where $\gamma_e$ is the the gyromagnetic ratio of the electronic spin ($\simeq 2.8 \mbox{ MHz/Gauss}$). Combined with the crystal-orientation-dependent Zeeman splitting and dynamical decoupling to narrow the linewidth, this resolution can go down to a few tens of nanometers\cite{dolde2013room, arai2015fourier}. 

\begin{figure*} [t]
\centering
\includegraphics[scale=0.42]{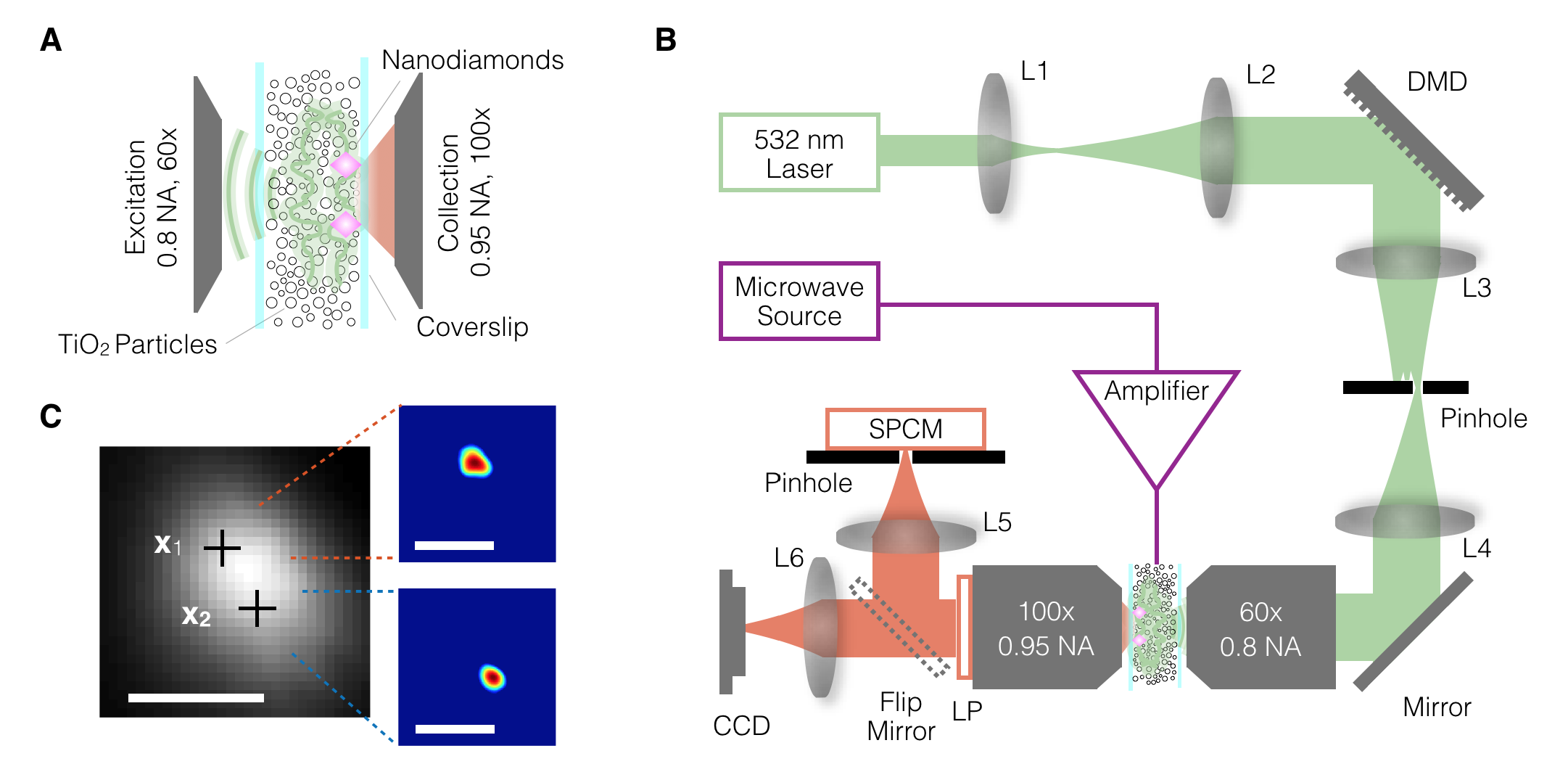}
\caption{
\textbf{Experimental Configuration.} 
(\textbf{A}) NV centers in subwavelength nanodiamonds (Diamond Nanotechnologies) are embedded in a complex medium consisting  of randomly distributed TiO$_2$ nanoparticles (Sigma Aldrich 718467). A green laser beam is delivered to the complex medium by a microscope objective (0.8 NA, 60x). An objective (0.95 NA, 100x) at the other side directly collects spin-dependent broadband red fluorescence from NV centers. The thickness of the complex medium is $\sim 7 \pm 2 \text{ }\upmu\text{m}$.
(\textbf{B}) Setup schematic: A DMD shapes the wavefront of the incident green laser and projects it onto the back aperture of excitation objective. The phase of each incident basis mode $\{ \mathbf{k} _n \}$ is controlled by groups of 24 by 24 DMD micro-mirrors. SPCM (CCD) counts (images) the red fluorescence collected by the collection objective. LP rejects the transmitted green laser, and a pinhole in front of SPCM blocks stray red fluorescence. A copper wire (diameter of 25 $\upmu\text{m}$) delivers the microwave signal to QRBs to modulate their spin ground state population. A permanent magnet (not shown) separates the magnetic resonance frequencies of the QRBs by orientation-dependent Zeeman splitting. (DMD: a digital micro-mirror device, SPCM: a single photon counting module, CCD: a charge-coupled device, LP: a long-pass optical filter with a cutoff wavelength of $650 \ \text{nm}$, and L1 - L6: lens)
(\textbf{C}) QRB fluorescence images for super-resolution focusing demonstration. $\mathbf{x}_1$ and $\mathbf{x}_2$ denote the QRB positions. Inset images are obtained using super-resolution focusing of our QRB-assisted wavefront shaping technique. (All scale bars = $0.61 \lambda / \text{NA}$ with NA = 0.8 and $\lambda = 532 \mbox{ nm}$) 
}
\end{figure*}

Figure 3 illustrates the experimental configuration for demonstrating QRB-assisted wavefront shaping. Our QRBs consist of  ensembles of NV centers (Fig. 1B) in nanodiamonds with a mean diameter of $50 \text{ nm}$. The QRBs are embedded in a complex medium consisting of randomly distributed TiO$_2$ nanoparticles with a mean diameter of $21 \text{ nm}$. The incident green laser light ($\lambda = 532 \mbox{ nm}$) is randomly scattered as it propagates through the medium. This scattering produces subwavelength spatial features on the incident laser light \cite{emiliani2003near,park2013subwavelength}, which excite the embedded QRBs. In particular, we demonstrate super-resolution focusing on two QRBs at $\mathbf{x}_1$ (QRB$_1$) and $\mathbf{x}_2$ (QRB$_2$) in Fig. 3C, where their separation $\abs{\mathbf{x}_1 - \mathbf{x}_2} = 186 \text{ nm}$ is far below the diffraction limit of our excitation objective lens, $406 \text{ nm}$ (Fig. S3). The QRB$_1$ (QRB$_2$) has the ESR frequency of $\nu_1 = 2.825 \text{ GHz}$ ($\nu_2 = 2.762 \text{ GHz}$), which corresponds to the electronic spin transition between $\ket{m_s = 0}$ and one of the Zeeman-split $\ket{m_s = \pm 1}$ of the ground spin triplet ($^3$A, Fig. 1B). Since $\nu_1$ and $\nu_2$ are well-separated ($\Delta \nu \simeq 63 \text{ MHz}$) compared to their resonance linewidths ($\delta \nu_1 = 5 \text{ MHz}$ and $\delta \nu_2 = 5.6 \text{ MHz}$), it is possible to individually drive the spin transition of each QRB.

In this study, we shape the incident wavefront with $793$ transverse Fourier basis modes $\{\mathbf{k}_n \}$, which cover the entire back aperture of the excitation objective. Resonant microwaves drive the spin transitions at $\nu_1$ and $\nu_2$ that produce the QRB-GS feedback, and the phase of $\{\mathbf{k}_n \}$ is iteratively adjusted to optimize the feedback signal (Fig. 2D and 2F). Figure 4A and 4B plot the results of the wavefront optimizations $W_{\nu_1}$ and $W_{\nu_2}$, respectively. For comparison, Fig. 4C shows the wavefront $W_\text{cl}$, obtained without the use of ESR (i.e. by optimizing only fluorescence feedback from QRBs). This fluorescence GS method\cite{vellekoop2008demixing,horstmeyer2015guidestar} focuses the interior optical field without achieving super-resolution. 

\begin{figure*}[t]
\centering
\includegraphics[scale=0.42]{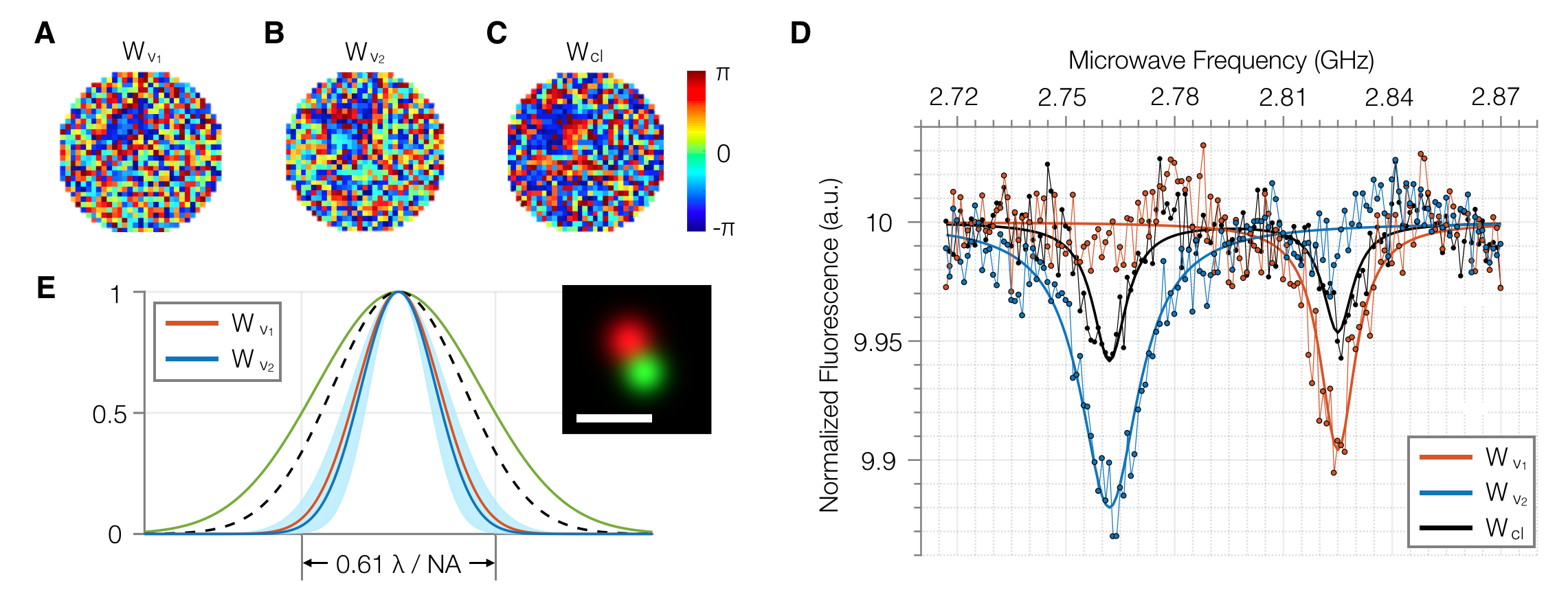}
\caption{
\textbf{Subwavelength optical focusing in a complex medium.} 
(\textbf{A}) and (\textbf{B}) The phase-only wavefronts $W_{\nu_1}$ and $W_{\nu_2}$,  determined by optimizing the QRB-GS feedback at $\nu_1$ and $\nu_2$, respectively. 
(\textbf{C}) The phase-only wavefront $W_\text{cl}$, obtained using the fluorescence GS method.
(\textbf{D}) ODMR spectra with $W_{\nu_1}$ (Red), $W_{\nu_2}$ (Blue), and $W_\text{cl}$ (Black) projection. 
(\textbf{E}) Spatial resolution of the subwavelength foci in the complex medium: The Red and Blue lines plot the estimated intensity shape of the subwavelength foci with $W_{\nu_1}$ and $W_{\nu_2}$ projection, respectively. The shaded area gives the estimation uncertainty. The Green line plots the point spread function (PSF) of the excitation objective (FWHM = $0.61\lambda / \mbox{NA}$ with $\lambda = 532 \mbox{ nm}$ and $\mbox{NA} = 0.8$), and the Black dashed line refers to the far-field limited PSF ($\mbox{NA} = 1$). Inset: reconstructed image of the subwavelength foci with $W_{\nu_1}$ (Red) and $W_{\nu_2}$ (Green). (Scale bar = $0.61 \lambda / \text{NA}$ with NA = 0.8 and $\lambda = 532 \mbox{ nm}$)
}
\end{figure*}

Projecting the wavefronts $W_{\nu_1}$ ($W_{\nu_2}$) forms a super-resolution optical focus at $\mathbf{x}_1$ ($\mathbf{x}_2$) in the complex medium. We can verify this super-resolution focusing by investigating optically-detectable-magnetic-resonance (ODMR) spectra. This is because (i) ODMR spectra exhibit resonances only of optically pumped QRBs, and (ii) QRB$_1$ and QRB$_2$ have distinguishable spectra. Figure 4D plots the ODMR spectra for this investigation. First, we project the wavefront $W_\text{cl}$ with the DMD (Fig. 3B), which produces the ODMR spectrum shown in the black line. This spectrum shows the resonances at $\nu_1$ and $\nu_2$ of both QRBs, as expected. By contrast, the only resonance of QRB$_1$ appears (the Red line) when we project $W_{\nu_1}$, which is obtained using the QRB-GS feedback with the spin transition at $\nu_1$. Alternatively, projecting $W_{\nu_2}$ reveals the resonance of QRB$_2$ (the Blue lines). This demonstration validates the ability of QRB-guided wavefront shaping to enable optical addressing of individual spots far below the diffraction limit. Note that the resonance linewidths are slightly broadened when the QRBs are excited by the targeted subwavelength foci, owing to the optically induced relaxation of ODMR\cite{dreau2011avoiding}. 

The ODMR spectra with subwavelength spin addressing enable us to estimate the spatial resolutions of the optical foci (Fig. 4E). We determine the peak-to-background intensity ratio of the focus (i.e. $I(\mathbf{x}_1) / I(\mathbf{x}_2)$ or vice versa) from the ODMR spectra (See Supplementary text S6 for details). Assuming the subwavelength focus features a Gaussian intensity envelop, the intensity ratio indicates that the super-resolution focus at $\mathbf{x}_1$ ($\mathbf{x}_2$) has a spatial resolution of $204 \text{ nm}$ ($184 \text{ nm}$). This achieved resolution is 2 (2.21) times smaller than our diffraction-limited resolution and 1.31 (1.45) times smaller than the far-field-limited one ($\mbox{NA} = 1$).

In conclusion, we introduced a quantum reference beacon (QRB) that enables super-resolution optical focusing within complex media. This QRB-GS approach uniquely provides, for the first time, sub-wavelength guidestar feedback inside a scattering medium by the use of spin coherence. Implementing our proposal with NV centers demonstrates clear super-resolution focusing capabilities inside of a complex medium. This QRB-assisted wavefront shaping opens up a range of applications. First, it can extend to quantum sensing based on NV centers to greater imaging depth and optical super-resolution. Second, it can be used to characterize the light propagation through a fiber for single-fiber endomicroscopy\cite{choi2012scanner}. Finally, our method could open up the way for subwavelength optical spin measurement\cite{rittweger2009sted,maurer2010far} of magnetic dipole-coupled quantum emitters\cite{dolde2013room}, which is essential for advanced quantum sensing\cite{giovannetti2011advances}, quantum error correction\cite{taminiau2014universal}, and room-temperature quantum computing \cite{yao2012scalable}. 

The authors would like to thank Noel H. Wan for his perspective on the manuscript. Donggyu Kim acknowledges financial support from Kwanjeong Educational Foundation. This research is supported in part by the Army Research office MURI biological transduction program and NSF Center for Integrated Quantum Materials. 

\def\theequation{S\arabic{equation}}
\def\thefigure{S\arabic{figure}}
\setcounter{equation}{0}
\setcounter{figure}{0} 

\section*{Supplementary Text}
\subsection*{S1. Electronic Spin Resonance of Quantum Reference Beacons}
The quantum reference beacon (QRB) consists of a solid-state quantum emitter with spin-dependent fluorescence. One example is a nitrogen-vacancy (NV) center in diamond, which has the electron spin magnetic sublevels $\ket{m_s = \pm 1}$ that are optically darker than $\ket{m_s = 0}$\cite{doherty2013nitrogen}. This spin-dependent fluorescence enables optical detection of magnetic resonance. The QRB-guidestar feedback is based on the optically detectable magnetic resonance (ODMR), in which we readout the optical fields at the resonance points inside of complex media. By driving the resonances of the QRBs, we individually modulate their spin populations below the optical diffraction-limited resolution, which produces the guidestar feedback for wavefront shaping. In our NV-based experiment, we drive the resonance between the bright $\ket{m_s = 0}$ and one of the dark $\ket{m_s = \pm 1}$ of NV centers. In the following, we represent the bright and dark spin state involved in the magnetic resonance as $\ket{0}$ and $\ket{1}$, respectively, and denote a spin state of a QRB with a spin density operator $\rho$. 

For our QRB-assisted wavefront shaping, the evolution of $\rho$ is described by the master equation 
\begin{equation}
\frac{d\rho}{dt} = \frac{1}{i\hbar} [H,\rho] + \left\{ \frac{d\rho}{dt} \right\}_\text{relax},
\end{equation} 
where $H$ is a simple two-level spin Hamiltonian that describes the relevant interaction of QRB with a microwave:
\begin{equation*}
H= \hbar \omega \dyad{1} + \hbar \Omega \cos (\omega_\text{mw}) (\dyad{0}{1}+\dyad{1}{0}). 
\end{equation*}
Here, the energy splitting $\hbar\omega$ includes the zero-field splitting $D_\text{gs} \simeq  2.87 \text{ GHz}$ and the electronic Zeeman splitting $\gamma_e B_{0z}$, where $\gamma_e = 2.8 \text{ MHz / Gauss}$ and $B_{0z}$ is a magnetic field along the symmetry axis of NV centers\cite{doherty2013nitrogen}. $\omega_\text{mw}$ is a microwave frequency that drives the magnetic resonance with the Rabi frequency $\Omega$.

The last term of the master equation Eq. (S1) represents the spin relaxation due to the interaction with the QRB's environment. This relaxation process includes the intrinsic spin relaxation from magnetic dipolar interactions with a spin bath. In addition, optical excitation induces spin relaxation through (i) spin polarization via intersystem crossing (ISC) followed by non-radiative decay, and (ii) the decoherence with scattered photons\cite{dreau2011avoiding}. Typical values of the intrinsic and optically induced relaxation rates for NV centers can be found in A. Dr\'eau \textit{et al.}\cite{dreau2011avoiding}

Modulating the spin state $\rho$ with a resonant microwave produces the guidestar feedback (the QRB-GS feedback). We denote the modulation through a quantum operator $E$ that maps an initial spin state $\rho$ to the modulated state $E(\rho)$. In our experiment, we modulate $\rho$ by continuous ESR (electronic spin resonance) spectroscopy, in which $\rho$ and $E(\rho)$ are the steady-steady solutions of Eq. (S1) under optical excitation. With the modulation, the change of the spin population $\Delta \sigma$ is
\begin{equation}
\Delta \sigma  = \frac{1}{2}  \text{tr} \big[ \sigma_z \{ \rho - E(\rho) \} \big], 
\end{equation}
where $\sigma_z$ is the Pauli-$z$ operator, and $\text{tr}[.]$ is the trace operator. For example, $\Delta \sigma  = 1/2$ for continuously-driven ESR of an initially polarized spin (i.e. $\rho = \dyad{0}$), and $\Delta \sigma = 1$ for an ideal spin flip with a microwave $\pi$-pulse. For the case that the microwave has a detuning $\delta$ from the spin resonance frequency, $\Delta \sigma$ is reduced by an ESR lineshape $g(\delta)$ with $g(0) = 1$: 
\begin{equation}
\frac{1}{2}  \text{tr} \big[ \sigma_z \{ \rho_\delta - E(\rho_\delta) \} \big] = g(\delta) \Delta \sigma.
\end{equation}
Here, $\rho_\delta$ refers to the spin state that is driven by the microwave with a detuning of $\delta$. Although details of the lineshape function depend on the dominant broadening mechanisms, we assume here that $g(\delta)$ is a Gaussian-shape function with a full-widht-at-half-maximum (FWHM) linewidth of $\delta \nu$. 

\subsection*{S2. The QRB-Guidestar Feedback}
In this section, we formulate the QRB-GS feedback in detail. As introduced in the main manuscript, $\pmb{\rho} =  \rho_1 \otimes \cdots \otimes \rho_M$ labels the initial spin state of QRBs at positions of $\mathbf{x}_1,  \cdots ,\mathbf{x}_M$ inside of the complex medium. An external magnetic field gradient separates individual resonance frequencies $\{\nu_m \}$ of $\{ \rho_m \}$ ($m = 1,2, \cdots, M$). $\{ E_m \}$ resonantly modulate the spin density operators $\{ \rho_m \}$ with lineshape functions $\{ g_m \}$ and  linewidths $\{ \delta\nu_m \}$. In principle, $\{\mathbf{x}_m \}$ could then be reconstructed from $\{\nu_m \}$ and knowledge of the external magnetic field. 

Optical fields inside of the complex medium are described by a transmission matrix $\vb{T}$. For example, the matrix element $t_{m,n}$ describes the optical field at $\{\mathbf{x}_m \}$ when the $n$th incident basis mode couples into the medium\cite{vellekoop2007focusing}, i.e., $\abs{t_{m,n}}^2$ is the optical intensity that excites the QRB at $\mathbf{x}_m $. The internal optical fields excite the embedded QRBs, which in turn emit spin-dependent broadband fluorescence. By driving the spin resonance of target QRBs, the spin-dependent fluorescence provides information of the transmission matrix elements.

Specifically, obtaining the QRB-GS feedback $S_{i,n}$ at $\mathbf{x}_i$ for the $n$th incident basis mode proceeds as follow. We first collect the fluorescence photons of the initial spin state $\pmb{\rho}$ with the basis mode. The photon numbers $N_n$ collected for unit integration time is:   
\begin{equation}
N_n[\pmb{\rho}] = \sum _m ^M \abs{t_{m,n}}^2 \{\gamma_0 \sigma_{00} ^m  + \gamma_1 \sigma_{11} ^m  \}.
\end{equation}
Here, $\gamma_0$ ($\gamma_1$) represents the collected spin-dependent photon numbers of $\ket{0}$ ($\ket{1}$) for unit excitation intensity. $\sigma_{00} ^m$ and $\sigma_{11} ^m$ account for the spin population (i.e. $\sigma_{00} ^m=\mel{0}{\rho_m}{0}$, $\sigma_{11} ^m = \mel{1}{\rho_m}{1}$) of the QRBs.  We assume that the photon collection are identical for all embedded QRBs. 

Next, we apply a microwave that is resonant to the target $i$th-QRB and repeat the fluorescence photon collection. As introduced in Eq. (S2) and (S3), the microwave operation, which is represented through a quantum operator $E_i$, modulates the spin population by $\Delta \sigma_i$ for the target QRB and by $g_m (\delta_m ^i) \Delta \sigma_m$ for the other `background' $m$th-QRB, where $\delta_m ^i = \nu_m - \nu_i$ (i.e. $\delta_i ^i = 0$). Then the collected photon number $N_n$ is 
\begin{align}
N_n[E_i(\pmb{\rho})] &= \abs{t_{i,n}}^2   \{ \gamma_0( \sigma_{00} ^i -  \Delta \sigma_i) + \gamma_1 (\sigma_{11} ^i + \Delta \sigma_i )   \} \nonumber \\
& \quad + \sum_{m \neq i} ^M \abs{t_{m,n}}^2 \{ \gamma_0( \sigma_{00} ^m - g_m(\delta_m ^i) \Delta \sigma_m) \nonumber \\
& \qquad \qquad \qquad  + \gamma_1 (\sigma_{11} ^m + g_m (\delta_m ^i) \Delta \sigma_m)    \}.
\end{align}
By subtracting Eq. (S5) from (S4), 
\begin{align} 
N_n[\pmb{\rho}] - N_n[E_i(\pmb{\rho})]  &= \abs{t_{i,n}}^2 \Delta \sigma_i (\gamma_0 - \gamma_1) \nonumber \\
& \quad + \sum_{m \neq i} ^M \abs{t_{m,n}}^2 g_m (\delta_m ^i) \Delta \sigma_m (\gamma_0 - \gamma_1).
\end{align}
If $\delta_m ^i \ge \delta \nu_m$ for all $m \neq i$, the contribution from the background QRBs is ignorable, reducing Eq. (S6) to the desirable QRB-GS feedback at $\mathbf{x}_i$: 
\begin{equation*}
S_{i,n}  = N_n[\pmb{\rho}] - N_n[E_i(\pmb{\rho})] = \abs{t_{i,n}}^2 \Delta \sigma_i (\gamma_0 - \gamma_1).
\end{equation*}

The condition $\delta_m ^i \ge \delta \nu_m$ determines our spatial resolution of the QRB-GS feedback, with an analogous to Rayleigh resolution limit in conventional optical microscopy. For a given external magnetic field gradient $dB/dx$, the ESR lineshape $g(\delta_m ^i)$ is translated to an effective point spread function (PSF) of the QRB-GS feedback. Thus, our spatial resolution $\Delta d_\text{QRB}$ is given by the FWHM  resolution of the effective PSF:
\begin{equation*}
\Delta d_\text{QRB} = \frac{\delta \nu}{\gamma_e (dB/dx)}.
\end{equation*}
Here, $\delta \nu$ is the mean ESR linewidth of QRBs (i.e. $\frac{1}{M} \sum_{m=1} ^M \delta \nu_m$). This spatial resolution can be improved by introducing assumptions such as $\Delta\sigma_ 1 = \Delta\sigma_2 = \cdots  =\Delta\sigma_M$. This resolution improvement depends on the accuracy of the assumptions and the signal-to-noise ratio of the measurements as in conventional optical microscopy \cite{goodman2005introduction}.

\subsection*{S3. Noise Estimation}
The noise in the QRB-GS feedback can be modeled with the photon shot noise of spin-dependent fluorescence. For the QRB-GS feedback
\begin{equation}
S_{i,n} = N_n[\pmb{\rho}] - N_n[E(\pmb{\rho})] = \abs{t_{i,n}}^2 \Delta \sigma_i (\gamma_0 - \gamma_1),
\end{equation}
we consider the photon shot noise in $N_n[\pmb{\rho}]$ and $N_n[E(\pmb{\rho})]$:
\begin{itemize}
\item{$\abs{t_{i,n}}\sqrt{  \gamma_0 \sigma_{00} ^i + \gamma_1 \sigma_{11} ^i  } $}
\item{$\abs{t_{i,n}}\sqrt{ \gamma_0 ( \sigma_{00} ^i - \Delta \sigma_i ) + \gamma_1 ( \sigma_{11} ^i + \Delta\sigma_i )   }$}
\item{$\sum_{m\neq i} ^M \abs{t_{m,n}} \sqrt{ \gamma_0 \sigma_{00} ^m  + \gamma_1 \sigma_{11} ^m}  $}
\item{$
\sum_{m\neq i} ^M \abs{t_{m,n}}  \\ 
\times\sqrt{ \gamma_0\{ \sigma_{00} ^m - g_m(\delta_m ^i) \Delta\sigma_m \} + \gamma_1 \{ \sigma_{11} ^m + g_m(\delta_m ^i)\Delta\sigma_m \}  }
$
}
\end{itemize}
Assuming that the complex medium is in a lossless waveguide whose cross-section area is $A$, the noise $N_{i,n}$ to signal Eq. (S7) ratio is
\begin{align*}
\frac{N_{i,n}}{S_{i,n}} 
&= \frac{
\sum_m \abs{t_{m,n}}  \sqrt{ 1 - \sigma_{11} ^m C} 
}
{
\abs{t_{i,n}}^2 \Delta\sigma_i \sqrt{\gamma_0}C
} \\
&\qquad\qquad \times \Bigg[	
1 + \sqrt{
1 -  \frac{\Delta \sigma_m g_m(\delta_m ^i)C }{1 - \sigma_{11}^m C}
}
\Bigg]
\\
&\simeq 
\frac{
\sum_m \abs{t_{m,n}}  \sqrt{ 1 - \sigma_{11} ^m C} 
}
{
\abs{t_{i,n}}^2 \Delta\sigma_i \sqrt{\gamma_0}C
} \\
&\qquad\qquad \times \Bigg[	
2 - 
  \frac{\Delta \sigma_m g_m(\delta_m ^i)C}{2(1 - \sigma_{11}^m C)}
\Bigg]\\
&\lesssim
\frac{
2 \sum_m \abs{t_{m,n}} 
}
{
\abs{t_{i,n}}^2 \Delta\sigma_i \sqrt{\gamma_0}C
}
\\
&\simeq 
\frac{
2 M \sqrt{(a/A)T} 
}
{
(a/A)T \Delta\sigma_i \sqrt{\gamma_0}C
}
\\
&=
\frac{2 M}{  
\sqrt{\gamma_0\tilde{a}T} \Delta\sigma_i C
}
. 
\end{align*}
Here, $\sum_{m=1} ^M \abs{t_{m,n}}^2 = (a/A) T \doteq \tilde{a}T$ where $a$ is the cross-sectional area of QRBs, $C = 1 - \gamma_1 / \gamma_0$, and $T$ is the total transmission of the complex medium. As expected, the noise-to-signal ratio approaches to zero for a longer integration.

\subsection*{S4. Four-Phase Method with the QRB-GS Feedback}
To access the phase of the transmission matrix element $t_{m,n}$, the incident Fourier basis modes $\{\mathbf{k}_n \}$ are encoded into the holographic illuminations by interfering themselves with the reference mode $u_\text{ref}$. Specifically, the holographic illumination of a incident basis mode $\mathbf{k}_n$ is represented by 
\begin{equation}
E^\text{in}_n  (\phi)= 1 + e^{\text{i}(\mathbf {k} _n  \cdot \mathbf{r} + \phi )},
\end{equation}
where we choose $ u_\text{ref} = 1 $, and $\phi$ is the phase of the basis mode relative to the reference mode. When $E^\text{in}_n (\phi)$ couples to the complex medium, the scattered optical field on the $i$th QRB is\cite{popoff2010measuring} 
\begin{align*}
	E_{i, n}^\text{out} ( \phi) &= \vb{T} E_n ^\text{in} =  t_{i, R} + t_{i,n}e^{\text{i}\phi}\\
		& =t_{i, R} \big(1 + \frac{t_{i,n}}{t_{i,R}}e^{\text{i}\phi} \big)  \\
		&= 1 + t_{i,n}e^{\text{i}\phi} ,
\end{align*}
where we substitute $t_{i,R} $ to $1$ without loss of generality. This leads the QRB-GS feedback $S_{i,n}(\phi)$:
\begin{align*}
S_{i,n}(\phi)
	&= \abs{E_{i,n} ^\text{out} (\phi)}^2 \times \Delta \sigma_i (\gamma_0 - \gamma_1)\\
	&= \big[1 + t_{i,n}^2 + 2\abs{t_{i,n}} \cos(\phi + \text{arg}(t_{i,n})) \big]  \\
	& \qquad \qquad \times \Delta \sigma_i (\gamma_0 - \gamma_1).
\end{align*}
By measuring the QRB-GS feedback with the four phase shifts $\phi = 0, \pi/2, \pi$, and $3\pi/2$, the phase of the matrix element is reconstructed by\cite{popoff2010measuring} 
\begin{align*}
	\text{arg}(t_{i,n}) &= \text{arg}\Big[\frac{S_{i,n}(0)-S_{i,n} (\pi)}{4} \\
	& \qquad \qquad \qquad + \text{i} \frac{S_{i,n} (3 \pi /2)-S_{i,n} (\pi/2)}{4} \Big].
\end{align*}

\subsection*{S5. Phase Readout with Continuously-Driven ESR}
We determine the phase of the transmission matrix element by sinusoidally modulating the optical excitation, as we sweep the phase of the incident basis modes relative to the reference mode. In the meantime, $\Delta \sigma$ with continuously-driven ESR depends on the optical excitation, since the steady-steady solutions $\rho$ and $E(\rho)$ of Eq. (S1) are a function of optical pumping \cite{dreau2011avoiding}. This dependence produces non-linearity of the four-phase shift measurement with the QRB-GS feedback. In this section, we show the small variation of $\Delta \sigma$ does not affect on the phase readout in our measurement up to the first order. 

For the holographic illumination $E^\text{in}_n=  1 + e^{\text{i}(\mathbf {k} _n  \cdot \mathbf{r} + \phi )}$, the optical excitation $I_{i,n} (\phi)$ on the $i$th QRB is 
\begin{equation*}
I_{i,n} (\phi) = \abs{E_{i,n} ^\text{out} (\phi)}^2 = 1 + t_{i,n}^2 + 2t_{i,n}\cos \big( \phi+ \theta_{i,n}   \big),
\end{equation*}
where we substitute $\text{arg}(t_{i,n})$ to $\theta_{i,n}$. We introduce the small variation of $\Delta \sigma$ up to the first order, while we modulate the phase $\phi$ of the holographic illumination:  
\begin{equation*}
\Delta \sigma _i (\phi) \simeq \Delta \sigma _i ^{(0)} + \Delta \sigma_i ^{(1)} (\phi)
\end{equation*}
where 
\begin{align*}
&\Delta \sigma_i ^{(0)}  = \eval{\Delta \sigma_i}_{I_{i,n}(0)} - I_{i,n}(0)\eval{\frac{d \Delta \sigma_i}{ dI }}_{I_{i,n}(0)} \\
&\Delta \sigma_i ^{(1)} (\phi) =  I_{i,n} (\phi)\eval{\frac{d \Delta \sigma_i}{ dI }}_{I_{i,n}(0)}. 
\end{align*}
The corresponding QRB-GS feedback is
\begin{align*}
S_{i,n} (\phi) & = I_{i,n}(\phi)  \Delta \sigma _i (\phi) (\gamma_0 - \gamma_1) \\
&=  I_{i,n}(\phi) [\Delta \sigma_i ^{(0)} + \Delta\sigma_i ^{(1)}(\phi)](\gamma_0 - \gamma_1) \\
&= S_{i,n}^{(0)}(\phi)+S_{i,n}^{(1)}(\phi),
\end{align*}
where
\begin{align*}
S_{i,n}^{(0)} (\phi)&=  (\gamma_0 - \gamma_1) \Delta \sigma_i ^{(0)}  I_{i,n} (\phi) \\
&\doteq \alpha_{i,n} ^{(0)} + \alpha_{i,n} ^{(1)} \cos (\phi + \theta_{i,n} ) \\
S_{i,n} ^{(1)} (\phi)&=  (\gamma_0 - \gamma_1) I_{i,n} (\phi)^2 \eval{\frac{d \Delta \sigma_i}{ dI }}_{I_{i,n}(0)}\\
	&\doteq\beta_{i,n} ^{(0)} + \beta_{i,n} ^{(1)} \cos ( \phi +\theta_{i,n} ) + \beta_{i,n} ^{(2)} \cos ^2 ( \phi + \theta_{i,n} ). 
\end{align*} 
Since $\cos^2 (\theta_{i,n} + \pi) =\cos^2 \theta_{i,n}  $ and $ \cos^2 (\theta_{i,n} + 3\pi/2) = \cos^2 (\theta_{i,n} + \pi/2) $, the nonlinear dependence in the four phase measurement is cancelled out, resulting in 
\begin{align*}
& \text{arg} \Big[\frac{S_{i,n} (0)-S_{i,n} (\pi)}{4} + \text{i} \frac{S_{i,n} (3 \pi/ 2)-S_{i,n} (\pi /2)}{4} \Big] \\ 
=& \text{arg} \Big[ \frac{\alpha_{i,n} ^{(1)} + \beta_{i,n} ^{(1)}}{2} \big( \cos\theta_{i,n}+ \text{i}\sin \theta_{i,n} \big) \Big]\\ 
=&  \text{arg}(t_{i,n}).
\end{align*}

\subsection*{S6. Estimation of Spatial Resolution}
Here, we describe how to estimate the spatial resolution of achieved subwavelength foci. Here, we assume the target QRB$_1$ and QRB$_2$ are point-like particles localized at $\mathbf{x}_1$ and $\mathbf{x}_2$, respectively. In continuously-driven ESR spectroscopy, the spin density operators $\rho_1$ and $\rho_2$ are optically polarized into $\dyad{0}$ when the microwave frequency $\nu_\text{off}$ is far off from their resonance frequencies, $\nu_1$ and $\nu_2$. By contrast, when the microwave is on resonances with $\nu_1$ and $\nu_2$, $\rho_1$ and $\rho_2$ become $(1-\Delta \sigma)\dyad{0} + \Delta \sigma\dyad{1}$ with $\Delta\sigma = 1/2$, provided that the QRBs are not optically saturated \cite{dreau2011avoiding}. 

In this analysis, we denote $I_1 ^{(1)}$ and $I_2 ^{(1)}$ ($I_1 ^{(2)}$ and $I_2 ^{(2)}$ ) as the optical excitation at $\mathbf{x}_1$ and $\mathbf{x}_2$ when the wavefront $W_{\nu_1}$ ($W_{\nu_2}$) is projected. $N^{(1)}(\nu)$ ($N^{(2)}(\nu)$) is the corresponding ODMR spectra with $W_{\nu_1}$ ($W_{\nu_2}$) projection. To estimate the spatial resolution $\Delta r^{(1)}$ of the subwavelength focus at $\mathbf{x}_1$ with $W_{\nu_1}$ projection, we consider the relations,
\begin{align}
N^{(1)} (\nu_\text{off})  &=  I_1 ^{(1)} \gamma_0 + I_2 ^{(1)} p \gamma_0  \\
N^{(1)}(\nu_1) &= I_1 ^{(1)} [\gamma_0 (1 - \Delta \sigma) + \gamma_1  \Delta \sigma)] + I_2 ^{(1)} p \gamma_0  \nonumber \\
& =  I_1 ^{(1)} (\gamma_0  - \frac{\Delta \gamma}{2} ) + I_2 ^{(1)} p \gamma_0 \\
N^{(1)}(\nu_2) &= I_1 ^{(1)} \gamma_0   + I_2 ^{(1)} p [\gamma_0 (1 - \Delta \sigma) + \gamma_1 \Delta \sigma] \nonumber\\
&= I_1 ^{(1)} \gamma_0   + I_2 ^{(1)} p (\gamma_0  - \frac{\Delta \gamma}{2} ).
\end{align}
Here, $\Delta \gamma = \gamma_0 - \gamma_1$, and the parameter $p$ takes account of the NV density difference between the two QRBs. $p$ can be determined from ODMR $N^{(\text{cl})}(\nu)$ under the diffraction-limited excitation, in which the optical excitations at $\mathbf{x}_1$ and $\mathbf{x}_2$ are approximately equal: 
\begin{align}
p =  \frac{N^{(\text{cl})}(\nu_\text{off}) - N^{(\text{cl})} (\nu_2)}{N^{(\text{cl})}(\nu_\text{off}) - N^{(\text{cl})} (\nu_1)}.
\end{align}
In our experiment, we determine $p$ with $W_\text{cl}$ projection.

\vspace{2.5mm}
From the ODMR spectra plotted in Fig. 4D, we obtain the ODMR at $\nu_\text{off}$, $\nu_1$, and $\nu_2$ by fitting to the Lorentzian lineshape function:
\begin{align*}
&N^{(1)}(\nu_1) / N^{(1)}(\nu_\text{off}) = 0.9902\ (0.9891, 0.9912)  \\
&N^{(1)}(\nu_2) / N^{(1)}(\nu_\text{off}) = 0.9988\  (0.9976, 0.9999) \\
&N^{(2)}(\nu_1) / N^{(2)}(\nu_\text{off}) = 0.9994\  (0.9985, 1.0004) \\
&N^{(2)}(\nu_2) / N^{(2)}(\nu_\text{off}) = 0.9878\ (0.9870, 0.9885) \\
&N^{(\text{cl})}(\nu_1) / N^{(\text{cl})}(\nu_\text{off}) = 0.9954\  (0.9943, 0.9965) \\
&N^{(\text{cl})}(\nu_2) / N^{(\text{cl})}(\nu_\text{off}) = 0.9942\  (0.9932, 0.9952), 
\end{align*}
where the values in the parenthesis represent $95\%$ confidence bound of the fitting. By inserting the fit values to Eq. (S9-S12), we found $I_1 ^{(1)}/I_2 ^{(1)}   \simeq 10.1$ and $p \simeq 1.261$. Similarly, $I_2 ^{(2)} / I_1 ^{(2)} \simeq17.1$.
Assuming subwavelength focus features a Gaussian intensity shape, its spatial resolution $\Delta r^{(1)}$ and $\Delta r^{(2)}$ are the FWHM of the intensity shapes:
\begin{align*}
\Delta r^{(1)} &=  2 \sqrt{\frac{  \ln 2}{ \ln (I_1 ^{(1)} / I_2 ^{(1)})}} \Delta x \simeq  204 \text{ nm} \\
\Delta r^{(2)} &=  2 \sqrt{\frac{  \ln 2}{ \ln (I_2 ^{(2)} / I_1 ^{(2)})}} \Delta x \simeq  184 \text{ nm},
\end{align*}
where $\Delta x = \abs{\mathbf{x}_1 - \mathbf{x}_2} =186 \text{ nm}$ (Fig. S3). We plot the uncertainties of the estimations in Fig. 4E.

\begin{figure*}[t]
\centering
\includegraphics[scale=0.42]{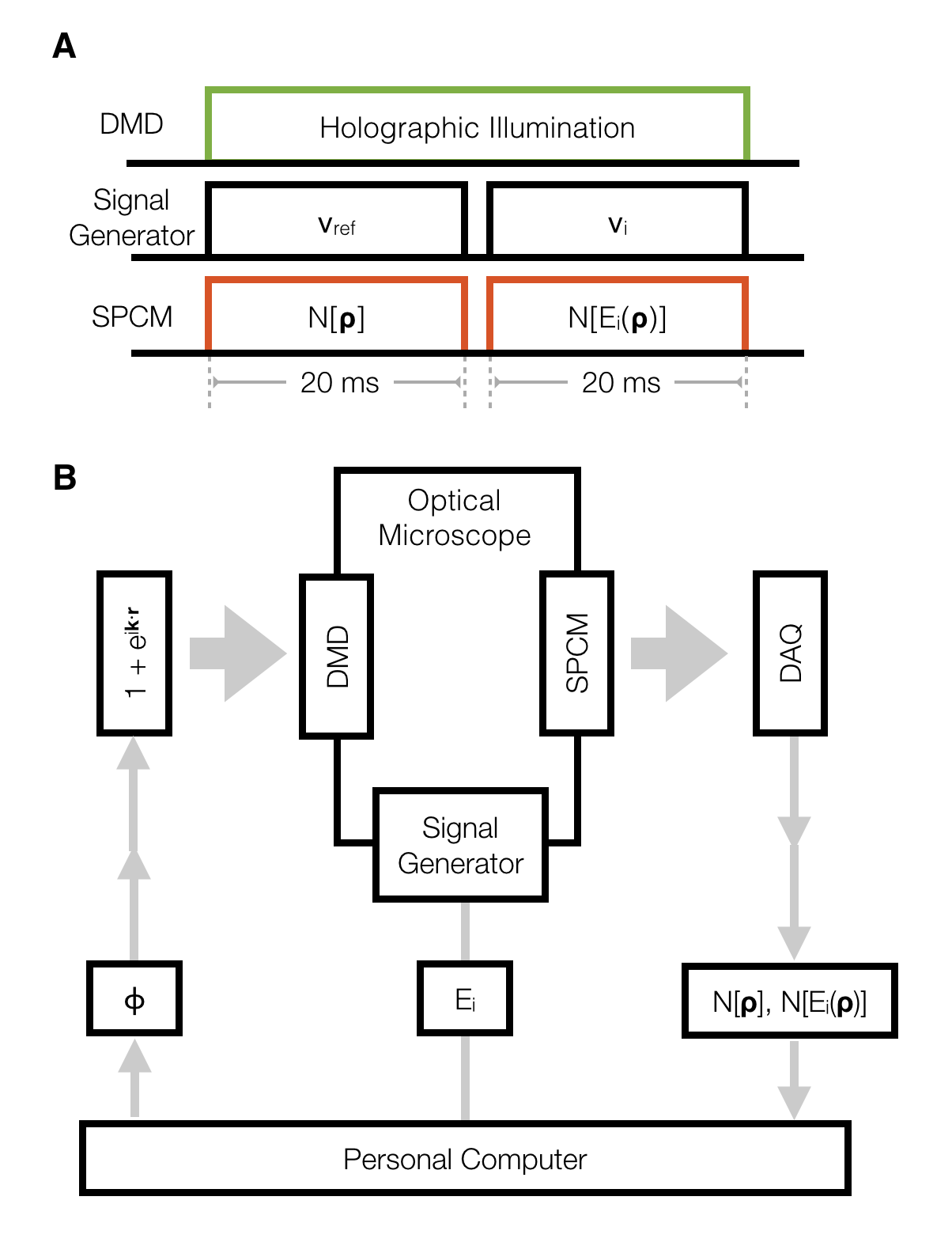}
\caption{
\textbf{The QRB-GS feedback measurement sequences}
(\textbf{A}) In our experiment, we modulate $\Delta \sigma_i$ of a target QRB with continuously-driven ESR. For a given holographic illumination, we continuously apply the microwave at the reference frequency of $\nu_\text{ref} = 2.5 \text{ GHz}$, which is far off from $\nu_1 = 2.825 \text{ GHz}$ and $\nu_2 = 2.762 \text{ GHz}$, for $20 \text{ ms}$, and at the the target resonance frequency $\nu_1$ or $\nu_2$ for another $20 \text{ ms}$. During the microwave operations, we simultaneously collect the spin-dependent fluorescence photon $N[\pmb{\rho}]$ and $N[E(\pmb{\rho})]$ with a single photon counting module. This unit sequence is repeated for 300 times per a holographic illumination. A digital clock pulse train from a DAQ synchronizes the microwave operations and the photon collections. 
(\textbf{B}) DMD in our optical microscope projects the holographic illumination $1 + e^{\text{i} \mathbf{k}\cdot \mathbf{r}}$ of the incident basis mode. Signal generator applies the quantum operator $E$ to produce $\Delta \sigma_i$. SPCM counts the spin-dependent fluorescence photons, and DAQ returns the fluorescence measurement $N[\pmb{\rho}]$, $N[E(\pmb{\rho})]$. From the measurements, personal computer determines the phase $\phi$ to be compensated on the incident basis mode. Updating the phase $\phi$ to DMD closes the iterative wavefront optimization cycle. DMD: digital micromirror device (D4100, Digital Innovations), SPCM: single photon counting module (SPCM-AQ4C, Excelitas), DAQ: Multi-functional data acquisition (NI-6343, National Instrument), Signal Generator (SME Rohde $\&$ Schwarz)
}
\end{figure*}

\begin{figure*}[t]
\centering
\includegraphics[scale=0.42]{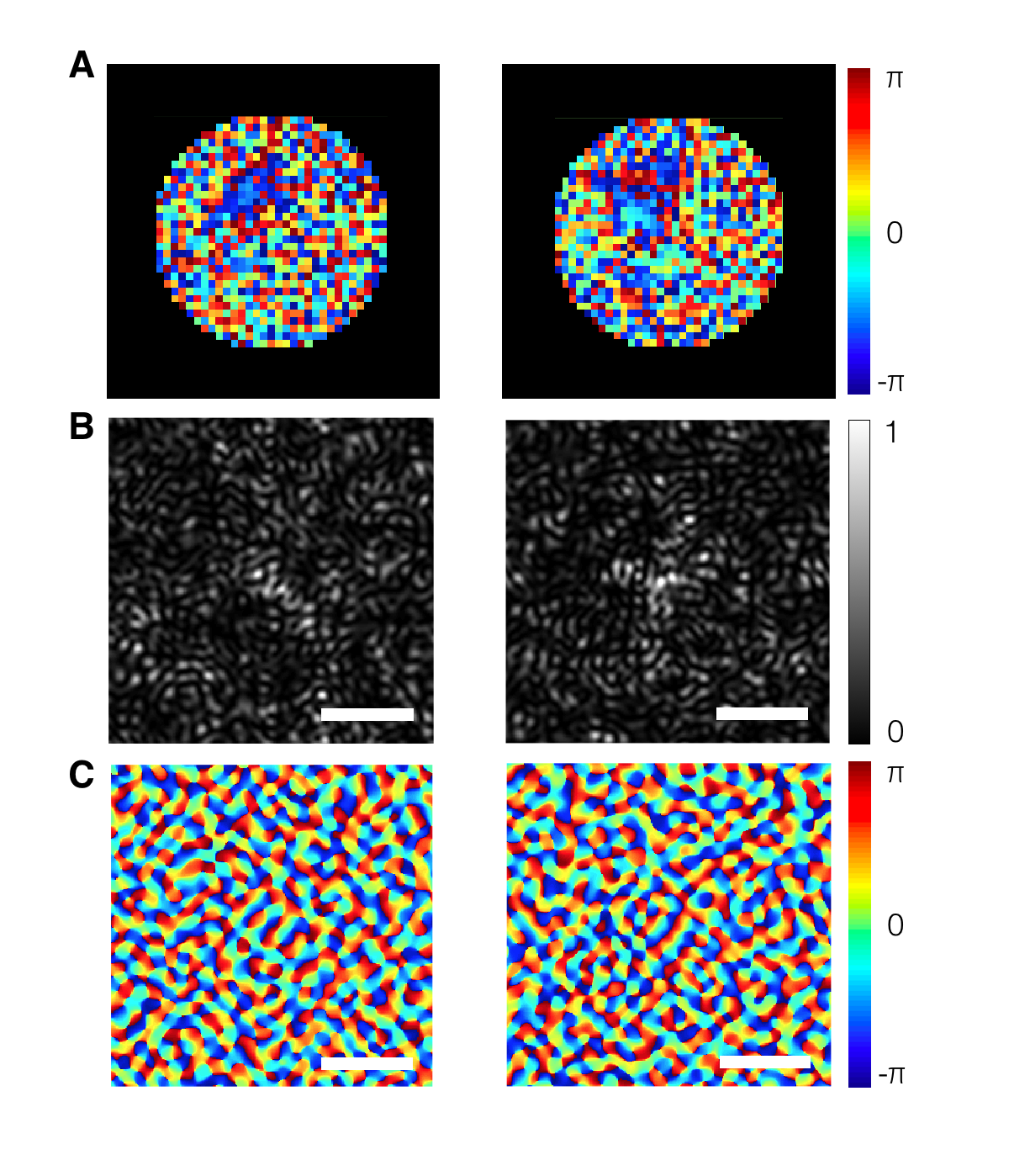}
\caption{
\textbf{Incident wavefront shaping with the QRB-GS feedback} (Scale bars = $3 \ \upmu\text{m}$). 
(\textbf{A}) The phase maps of 793 incident basis modes, determined by the QRB-GS feedback. The DMD projects the phase maps into the back aperture of the excitation objective lens. Left (Right) plot is the result of the iterative optimization with the QRB-GS feedback at $\nu_1$ ($\nu_2$). 
(\textbf{B}) and (\textbf{C}), The intensity and phase map of the incident wavefront on the complex medium, respectively.  These maps are obtained by the Fourier transform of the phase maps plotted in (\textbf{A}). 
}
\end{figure*}

\begin{figure*}[b]
\centering
\includegraphics[scale=0.42]{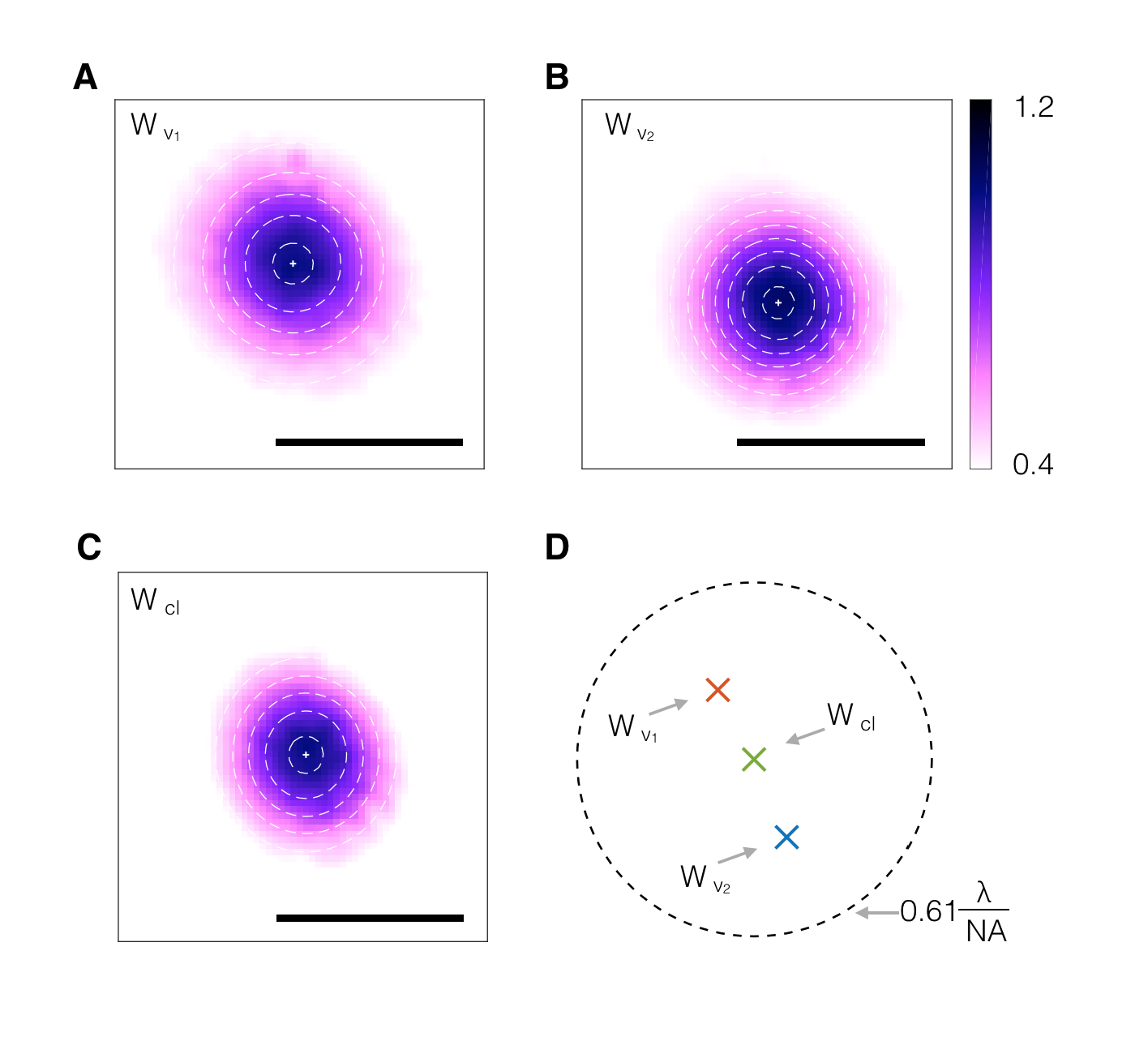}
\caption{
\textbf{Sub-diffraction localization of target QRBs}
(Scale bar = $1.22\lambda / \text{NA} \simeq 810 \text{ nm}$).
(\textbf{A}) and (\textbf{B}) We individually excite QRB$_1$ and QRB$_2$ with subwavelength optical focus through a scattering medium. The central position $\mathbf{x}_1$ and $\mathbf{x}_2$ of QRB$_1$ and QRB$_2$ are localized by fitting the recorded fluorescence images into two-dimensional Gaussian functions. 
(\textbf{C}) The central position with the $W_\text{cl}$ projection for comparison. 
(\textbf{D}) Merged positions from (\textbf{A}), (\textbf{B}), and (\textbf{C}). The dashed circle guides the diffraction-limited resolution of the excitation microscope objective (NA = 0.8, $\lambda = 532 \text{ nm}$).
}
\end{figure*}

\end{document}